\documentclass[aps]{revtex4}

\usepackage{graphicx}
\usepackage{dcolumn}
\usepackage{bm}

\begin{document}

\title{Robust Half-Metallic Character and Large Oxygen Magnetism
in a Perovskite Cuprate}

\author{Xiangang Wan\(^{1,2}\), Masanori Kohno\(^{1}\), and Xiao
Hu\(^{1}\)}

\affiliation{ \(^{1}\)Computational Materials Science Center,
National Institute for Materials Science, Tsukuba 305-0047, Japan
\\ \(^{2}\)National Laboratory of Solid State Microstructures and
Department of Physics, Nanjing University, Nanjing 210093, China }

\begin{abstract}
The new perovskite cuprate material
Sr$_{8}$CaRe$_{3}$Cu$_{4}$O$_{24}$, which behaves
ferrimagnetically and shows an unusually high Curie temperature
($T_c \sim$ 440 K), is found from density-functional theory
calculation to display several surprising properties after hole
doping or chemical substitution: (1) Half metal (HM) is realized
by replacing Re with W or Mo while $T_c$ remains high; (2)
hole-doped Sr$_{8}$CaRe$_{3}$Cu$_{4}$O$_{24}$ is also HM with high
$T_c$. Moreover, we find that the O atoms will carry a large
magnetic moment after hole doping, which is in sharp contrast with
the generally accepted concept that magnetism in solid requires
partially filled shells of $d$ or $f$ electrons in cations. The
material Sr$_8$CaRe$_3$Cu$_4$O$_{24}$ is therefore expected to
provide a very useful platform for material design and
development.
\end{abstract}


\date{\today}

\maketitle

\vspace{1cm}

Spin-based electronics, or spintronics, is currently a rather hot
topic, which offers opportunities for new generation of
multifunctional devices combining traditional charge-based
microelectronic with spin-dependent effects \cite{H-M-1}. An ideal
component for such devices is half-metallic ferromagnet, which is
metallic for one spin channel and insulating for the other
\cite{HM}. The search for HM materials with high $T_{c}$ is
intense in recent years \cite{H-M-2,H-M-3}. Up to present, many
works focus on the perovskite transition-metal oxides, in
particular the Mn-based and Fe-based compounds \cite{H-M-4,H-M-5}.
To find other new HM compounds with $T_c$ above room temperature
is of both fundamental and technological importance.

Since ferromagnetic (FM) cuprates are very rare, and usually show
rather low $T_c$, they have not been considered seriously as
candidates of useful magnetic material. Recently a new perovskite
cuprate material Sr$_{8}$CaRe$_{3}$Cu$_{4}$O$_{24}$ was
synthesized \cite{Takayama-Muromachi}. This cuprate forms the
cubic perovskite $AB$O$_{3}$ structure with space group $Pm-3m$.
The $A$-sites are occupied by Sr exclusively, Ca, Re and Cu are
located at the $B$-sites in an ordered way, and the unit cell
contains eight perovskite-like blocks. According to the symmetry,
the 24 O atoms in unit cell can be sorted into three kinds, O1, O2
and O3, the four Cu atoms are sorted into two kinds, one Cu1 and
three Cu2 \cite{Takayama-Muromachi}. This material has some unique
physical properties: it is similar in structure to the cuprates
showing high-temperature superconductivity upon carrier doping,
while possesses a net magnetization; its $T_c$ is over room
temperature ($\sim$ 440K), much higher than other known cuprates.
Unfortunately, this compound is an insulator, and cannot show any
magnetoresistance. Therefore, it is interesting to find a way to
make this compound electrically transportable meanwhile keeping
its high $T_c$. There are two well-known methods to tailor the
properties of perovskite materials: $B$-site substitution and
carrier doping. The former modifies the band structure while the
latter mainly shifts the Fermi level. As revealed in this Letter,
this cuprate shows several peculiar properties after appropriate
$B$-site substitution and carrier doping.

First let us consider the $B$-site element substitution.  A
valuable insight for possible material developing is available
starting from the band structure of Sr$_8$CaRe$_3$Cu$_4$O$_{24}$
\cite{wan-xg}. In Fig. 1 we show schematically the minority-spin
$e_g$ band of Cu1 and $e_g$-like band of Cu2, and the $p$ band of
O2, noting that other bands have small contributions to states
near the Fermi level and can be neglected for a qualitative
argument. The minority-spin $e_g$ bands of Cu1 and Cu2 are
distributed on the two sides of the Fermi level, with a large
interval between the two regions as shown in Fig.1 (a). Due to the
Jahn-Teller distortion caused by Re (the (010) plane is shown in
Fig. 1(a)), the degeneracy of the $e_g$ band of Cu2 is lifted and
the $d_{x^2-y^2}$ band becomes fully occupied. If one substitutes
Re by some other 5$d$ or 4$d$ elements with smaller
electronegativity, the Jahn-Teller distortion in the oxygen
octahedron centered at Cu2 should decrease. The splitting of $e_g$
band of Cu2 should decrease significantly, and as the result, it
may cross the Fermi level as shown in Fig.1 (b). Tungsten (W) and
molybdenum (Mo), the elements at the left column of Re in the
Periodic Table, can be good candidates for this purpose. Since
substituting Re with other elements will not affect much the bands
of Cu1, the spin-up channel therefore remains insulating. Same as
Sr$_8$CaRe$_3$Cu$_4$O$_{24}$ \cite{wan-xg}, the spin-up $e_g$ band
of Cu1 and spin-down $d_{3z^2-r^2}$ band of Cu2 are partially
occupied, which supports the strong \(pd\sigma\) hybridization.
Therefore, it is expected that substituting Re by W or Mo will
result in a HM with high $T_c$.
The above idea is verified by
accurate density-functional theory calculations as following.

Calculations have been performed with the WIEN2K
package\cite{Blaha}, which uses a full-potential, all electron
APW+lo method that allows one to carry out total energy
calculations of high precision. We adopt the standard generalized
gradient approximation (GGA) \cite{Perdew}, use the LSDA+U method
\cite{Anisimov} to treat the electron-electron interaction effects
with $U$ = 10 eV and $J$ = 1.20 eV for the {\it d} orbital of Cu
\cite{U}. All the results shown in the following are based on the
fully optimized structure. We have also confirmed that the main
results are robust and do not change with the values of
parameters.

For Sr$_8$CaW$_3$Cu$_4$O$_{24}$ we calculate both ferrimagnetic
(FiM) and FM configurations for Cu1 and Cu2 moments. The spin-up
$e_g$ band of Cu1 and spin-down $d_{3z^2-r^2}$ band of Cu2 are
partially occupied as shown in Fig. 2, while the majority-spin
bands are fully occupied. Therefore, the magnetic properties of
Sr$_8$CaW$_3$Cu$_4$O$_{24}$ is similar to that of
Sr$_8$CaRe$_3$Cu$_4$O$_{24}$ \cite{wan-xg}. The ground state is
FiM; the magnetic moments are carried mainly by the Cu ions; the
total moment per unit cell and the moments at Cu1 and Cu2 are
4.00, -0.89, and 1.02 $\mu_B$ respectively. The orbital ordering
in the $d$ bands of Cu1 and Cu2 revealed in Ref. \cite{wan-xg}
still exists, which induces a large hybridization with the $p_z$
orbital of O2, corresponding to the strong $pd\sigma$ bond, and
produces a strong superexchange interaction between the magnetic
moments at Cu1 and Cu2. As the result, relative to the FM
configuration, the FiM configuration gains 0.046 Ry per formula
unit. Comparing with the same energy difference of 0.036 Ry for
Sr$_8$CaRe$_3$Cu$_4$O$_{24}$, we predict that the $T_c$ of
Sr$_8$CaW$_3$Cu$_4$O$_{24}$ should be higher than 440 K. After
substituting Re by W, the Jahn-Teller distortion in the oxygen
octahedron centered at Cu2 becomes rather small
\cite{bond-length}, and as shown in Fig. 2 the splitting of $e_g$
orbitals of Cu2 is reduced from that of
Sr$_8$CaRe$_3$Cu$_4$O$_{24}$ (see Fig. 3 of Ref. \cite{wan-xg}).
Since the spin-down e$_g$-like bands of Cu2 cross the Fermi level,
while the spin-up channel is still insulating as shown in Fig. 2,
the new material Sr$_8$CaW$_3$Cu$_4$O$_{24}$ is a HM.

Since Mo and W lie at the same column of periodic table, the
properties of Sr$_8$CaMo$_3$Cu$_4$O$_{24}$ are similar to that of
Sr$_8$CaW$_3$Cu$_4$O$_{24}$. The energy difference between FiM and
FM configuration is also large, so one can expect that
Sr$_8$CaMo$_3$Cu$_4$O$_{24}$ still has high $T_c$. As expected,
the Jahn-Teller distortion in the oxygen octahedron centered at
Cu2 is also rather small. As a result, the spin-down channel is
metallic and the spin-up one is insulating as shown in Fig. 3.
Therefore, Sr$_8$CaMo$_3$Cu$_4$O$_{24}$ is also a HM with high
$T_c$.

Now let us turn to the effect of hole doping. It is interesting to
note that, as shown schematically in Fig.1(c), in
Sr$_8$CaRe$_3$Cu$_4$O$_{24}$ the $p_x$+$p_y$ band of O2 is rather
narrow and just below the Fermi level \cite{wan-data}. From the
view of a rigid band model, doping hole shifts the Fermi level
downward and causes it to cross the narrow $p_x$+$p_y$ band of O2.
Therefore, we expect that this compound may display some peculiar
properties related to O atoms after hole doping. In order to keep
the band structure, hole doping is realized for example by
$A$-site element substitution. Since the $A$-site substitution can
be treated successfully by the virtual-crystal approximation
(VCA)\cite{VC}, we employ it in the present study to calculate the
electronic and magnetic properties of
(Sr$_{1-x}A_x$)$_{8}$CaRe$_{3}$Cu$_{4}$O$_{24}$ with $A$ stands
for a univalent cation. It is found that, different from replacing
Re with W or Mo, doping hole has small effect on the geometric
structure. Since the states just below the Fermi level are almost
entirely due to the $p_x+p_y$ band of O2, the $d$ bands of Cu1 and
Cu2 are basically unaffected by the hole doping. Consequently,
same as Sr$_8$CaRe$_3$Cu$_4$O$_{24}$, the ground state is FiM, and
the energy difference between FM and FiM configuration is still
large, which implies a high $T_c$ of
(Sr$_{1-x}A_x$)$_8$CaRe$_3$Cu$_4$O$_{24}$. The magnetic moments at
Cu1 and Cu2 are almost independent of the hole doping as shown in
Table I. However, in striking contrast to the parent compound, O2
carries a large magnetic moment after hole doping. At doping
concentration $x$ = 0.2, the magnetic moment at O2 reaches 0.25
$\mu_B$, the largest value in literatures \cite{O-moment}, and
increases further with $x$ as shown in Table I. This phenomenon is
very peculiar since it is generally believed that large magnetic
moment is carried only by the transition-metal ions with unfilled
$d$ or $f$ bands.

In order to clarify the origin of large magnetic moment at O2, we
show the density of states (DOS) of
(Sr$_{0.8}A_{0.2}$)$_8$CaRe$_3$Cu$_4$O$_{24}$ in Fig. 4. (The DOS
of other doping $x$ is similar to that of $x$ = 0.2.) The $p_z$
state of O2, which strongly overlaps with the $d$ states of Cu1
and Cu2, forms a wide band. As a result, the exchange splitting of
the $p_z$ state of O2 is very small, which has only small
contribution to the magnetic moment. So we can conclude that the
magnetic moment at O is not induced by the magnetic ions Cu,
completely different from other known materials with magnetic
moment at O atoms. It is intriguing to note that the spin-up
$p_x$+$p_y$ band of O2 appears in the energy window centered at
$\sim$ -1.2 eV and are fully occupied, while the spin-down states
are obviously higher in energy than the spin-up one and cross the
Fermi level. This clearly indicates that the exchange splitting of
$p_x$+$p_y$ state of O2 is rather large, and the magnetic moment
at O2 is carried mainly by the $p_x$+$p_y$ state. The mechanism of
this exchange splitting can be attributed to the Stoner
instability as shown schematically in Fig.1 (d), namely when the
Fermi level crosses the narrow $p_x$+$p_y$ band of O2 upon hole
doping, the holes fully polarize in order to reduce the total
energy. Furthermore, the fully polarized holes lead this compound
to a HM as can be seen from Fig. 4.

Having clarified where the magnetic moment of O comes from, we try
to understand the reason why it is rare in oxides and thus needs
fine tuning by material design. As discussed above, there are two
essential factors for this phenomenon: the $p$ band of O should be
narrow and located nearby the Fermi level.  It is well known that
for perovskite $AB$O$_3$ compound, the occupied $d$ state shifts
to lower energy and the weight of the O 2$p$ states increases at
the top of valence band as the atomic number of the
transition-metal $B$ increases from 22 (Ti) to 29 (Cu)
\cite{trend-1}. This is the reason why O has the main contribution
to the top of valence band of Sr$_8$CaRe$_3$Cu$_4$O$_{24}$. In
contrast, in manganites and many other oxides the state just below
the Fermi level comes from the $d$ band of Mn mixing with the
$p$ band of O. 
Since Sr$_8$CaRe$_3$Cu$_4$O$_{24}$ has the cubic symmetry, the
angle of Cu1-O2-Cu2 is 180$^0$, the $e_g$ orbital of Cu1 and the
$e_g$-like orbital of Cu2 cannot hybridize with $p_x$+$p_y$
orbital of O2 due to symmetry. On the other hand, the $t_{2g}$
orbitals of Cu1 and Cu2 are fully occupied, and located at low
energy region due to the large Coulomb interaction, therefore the
hybridization between these orbitals with the $p_x$+$p_y$ orbitals
of O2 is rather small. Consequently, the $p_x$+$p_y$ band of O2 is
very narrow, and the DOS is rather high. In contrast, in most
perovskite $AB$O$_3$ materials, the $B$O$_{6}$ octahedron tilts
with the angle of $B$-O-$B$ smaller than 180$^0$, consequently the
$e_g$ orbital of $B$ can overlap with the $p_x$+$p_y$ orbitals of
O, and makes the band of the latter wide, typically of band width
of 5-7eV.  As it is larger than the intra-atomic exchange
strength, the O is non-magnetic in most oxides. Since the radii of
univalent cation $A$, such as potassium (K), is slightly larger
than that of Sr ion, hole doping into Sr$_8$CaRe$_3$Cu$_4$O$_{24}$
through substituting Sr by K will slightly enlarge the tolerance
factor. Therefore, the CuO$_{6}$ octahedron is not expected to
tilt upon such hole doping treatment.

Large magnetic moments at O2 upon hole doping are also observed
when the LSDA scheme is adopted, which tends to underestimate the
magnetic moment. Therefore, the present result is robust with
respect to $U$, and is likely to be observed experimentally.  Our
theoretical prediction may fail in several exceptional cases,
where the system happens to be close to a quantum critical point,
or an instability to the triplet superconductivity.  In any case,
interesting physical properties are expected, and deserve careful
experimental investigations.

In summary, using a density-functional calculation we investigate
the effect of substitution and hole doping on the material
Sr$_8$CaRe$_3$Cu$_4$O$_{24}$. Our study reveals that a HM with
high $T_c$ can be realized by replacing Re with W or Mo, and that
hole doping to the material will also result in a HM with high
$T_c$ and unusually large magnetic moments at O atoms due to the
Stoner instability.  The material Sr$_8$CaRe$_3$Cu$_4$O$_{24}$ is
therefore expected to provide a very useful platform for material
design and development, and potential products of the research in
this direction can be very useful for spintronics applications.

We are grateful to Dr. E. Takayama-Muromachi, and Dr. M. Isobe for discussions.
Calculations have been performed on Numerical Materials Simulator
(HITACHI SR11000) at the Computational Materials Science Center,
National Institute for Materials Science.
This work was partially supported by Japan Society for the Promotion of
Science (Grant-in-Aid for Scientific Research (C) No. 15540355).

\newpage

\begin{table}
\caption{Calculated total magnetic moment per unit cell
\(\mu_{tot}\), magnetic moments inside the muffin-tin spheres of Cu1,
Cu2 and O2 in units of \(\mu_{\rm B}\). } \vskip 5mm
\begin{tabular}{lccccc}
                Doping  & \(\mu_{\rm tot}\) & Cu1 & Cu2  &O2&
              \\
\hline 0        & -1.00& 1.08 &-0.81 & 0.07 \\
\hline 0.05    & -0.60& 1.06 &-0.81 & 0.12 \\
\hline 0.1     & -0.20& 1.04 &-0.80 & 0.16 \\
\hline 0.15    &  0.20& 1.05 &-0.81 & 0.20 \\
\hline 0.2      &  0.60& 1.04 &-0.80 & 0.25 \\
\hline 0.25    &  1.00& 1.04 &-0.80 & 0.29 \\
\hline 0.75    &  5.00& 1.07 &-0.79 & 0.59 \\
\hline 1    &  7.00& 1.08 &-0.80 & 0.70 \\
\hline
\end{tabular}
\end{table}

\vspace{5cm}

\noindent {\bf Figure Captions}

\noindent Fig. 1: (a) Schematic picture for the density of states
(DOS) of minority spin $e_g$ orbitals of Cu1 and Cu2, and the
partial structure of (010) plane in Sr$_8$CaRe$_3$Cu$_4$O$_{24}$.
(b) Schematic picture for the DOS of minority spin $e_g$ orbitals
of Cu1 and Cu2, and the partial structure of (010) plane expected
for Sr$_8$CaW$_3$Cu$_4$O$_{24}$ and Sr$_8$CaMo$_3$Cu$_4$O$_{24}$.
(c) Schematic picture for the DOS of $p$ orbital of O2 in
Sr$_8$CaRe$_3$Cu$_4$O$_{24}$. (d) Schematic picture for the DOS of
$p$ orbital of O2 expected for hole-doped
Sr$_8$CaRe$_3$Cu$_4$O$_{24}$. The dotted lines denote the Fermi
level.

\vspace{1cm}

\noindent Fig. 2: Total DOS (top) and partial DOS of the minority
spin (spin-up) $d$ orbitals of Cu1 (middle) and the minority spin
(spin-down) $d$ orbitals of Cu2 (bottom) of
Sr$_8$CaW$_3$Cu$_4$O$_{24}$ with the Fermi level set at zero.

\vspace{1cm}

\noindent Fig. 3: Total DOS of Sr$_8$CaMo$_3$Cu$_4$O$_{24}$ with
the Fermi level set at zero.

\vspace{1cm}

\noindent Fig. 4: DOS of
(Sr$_{0.8}A_{0.2}$)$_8$CaRe$_3$Cu$_4$O$_{24}$ with the Fermi level
set at zero.


\newpage
\begin{thebibliography}{9}

\bibitem{H-M-1} S.A. Wolf {\it et al.}, Science {\bf 294},
1488 (2001); I. {\u Z}uti{\' c}, J. Fabian and S. D. Sarma, Rev.
Mod. Phys. {\bf 76}, 323 (2004).

\bibitem{HM} W.E. Pickett and J.S. Moodera, Phys. Today {\bf 54},
39 (2001).

\bibitem{H-M-2} F.J. Jedema, A.T. Filip and B.J. van Wees, Nature
(London) {\bf 410}, 345 (2001); P. Sharma {\it et al.}, Nature
Materials {\bf 2}, 673 (2003); S.M. Watts {\it et al.}, Phys. Rev.
B {\bf 61}, 9621 (2000); S. Soeya {\it et al.}, Appl. Phys. Lett.
{\bf 80}, 823 (2002).

\bibitem{H-M-3} W.H. Xie {\it et al}, Phys. Rev. Lett. {\bf 91},
037204 (2003); A. Continenza {\it et al}, Phys. Rev. B {\bf 64},
085204 (2001); S. Sanvito and N. A. Hill, Phys. Rev. B {\bf 62},
15553, (2000).

\bibitem{H-M-4} See for example, {\it Colossal Magnetoresistive
Oxides}, ed. Y. Tokura, (Gordon $\&$ Breach Publishers, 1999).

\bibitem{H-M-5} K.I. Kobayashi, {\it et al.} Nature (London) {\bf
395}, 677 (1998).

\bibitem{Takayama-Muromachi} E. Takayama-Muromachi {\it et al.},
J. Solid State Chem. {\bf 175}, 366 (2003).

\bibitem{wan-xg} X. Wan, M. Kohno, and X. Hu, Phys. Rev. Lett.
{\bf 94}, 087205 (2005).

\bibitem{Blaha} P. Blaha {\it et al.}, {\bf WIEN2K}, An Augmented
Plane Wave + Local Orbitals Program for Calculating Crystal
Properties (Karlheinz Schwarz, Tech. Universit$\ddot a$t Wien,
Austria, 2001). ISBN 3-9501031-1-2
%
%
\bibitem{Perdew} J.P. Perdew, K. Burke, and M. Ernzerhof,
Phys. Rev. Lett. {\bf 77}, 3865 (1996).
%
\bibitem{Anisimov} V.I. Anisimov {\it et al.}, Phys. Rev. B {\bf 48},
16929 (1993).
%
\bibitem{U} V.I. Anisimov {\it et al.}, Phys. Rev. B {\bf 66},
100502(R) (2002).
%
\bibitem{bond-length} The Cu2-O2 and Cu2-O3 bond in optimized
SrCaW$_3$Cu$_4$O$_{24}$ is 2.00 and 2.03 $\AA$, respectively. In
contract, the Cu2-O2 and Cu2-O3 bond in SrCaRe$_3$Cu$_4$O$_{24}$
is 2.00 and 2.12 $\AA$, respectively.
%
%
\bibitem{wan-data} X. Wan, M. Kohno and X. Hu, unpublished.
%
\bibitem{VC} W.E. Pickett and D.J. Singh, Phys. Rev. B
{\bf 55}, R8642 (1997); P. Nov{\' a}k, {\it et al.}, Phys. Rev. B
{\bf 63}, 235114 (2001).
%
\bibitem{O-moment} R. Weht and W.E. Pickett, Phys. Rev. Lett. {\bf
81}, 2502 (1998).
%
\bibitem{trend-1} J. Zaanen, G.A. Sawatzky snd J.W. Allen,
Phys. Rev. Lett. {\bf 55}, 418 (1985).

%

\end{thebibliography}
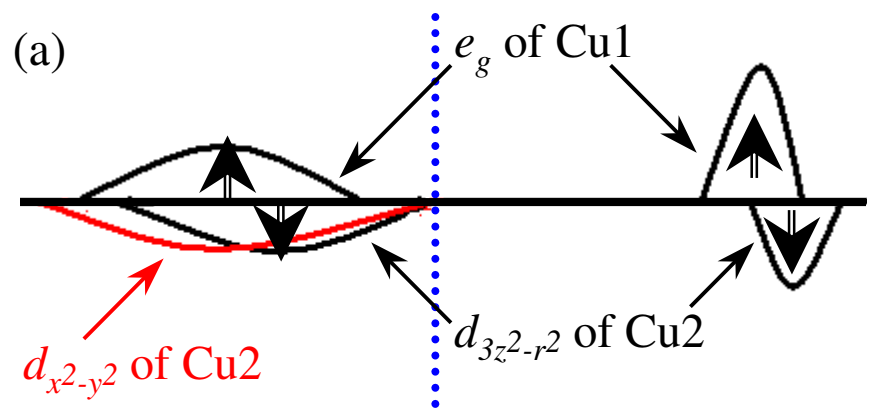
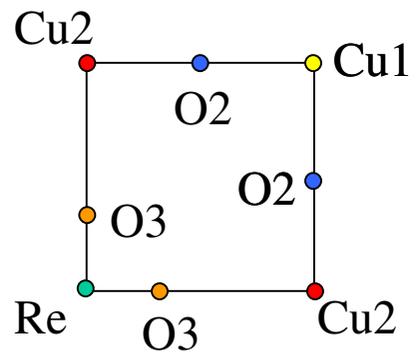
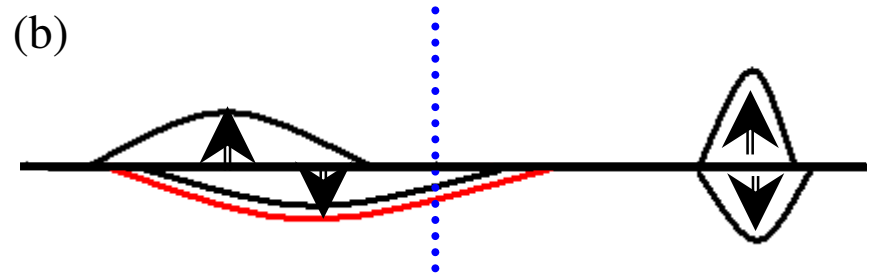
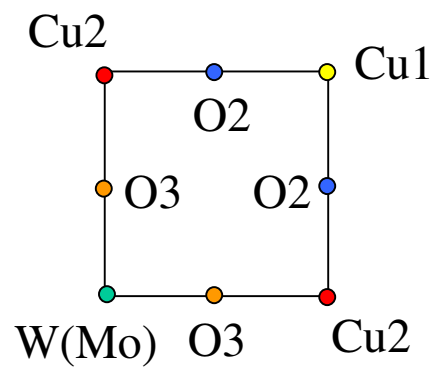
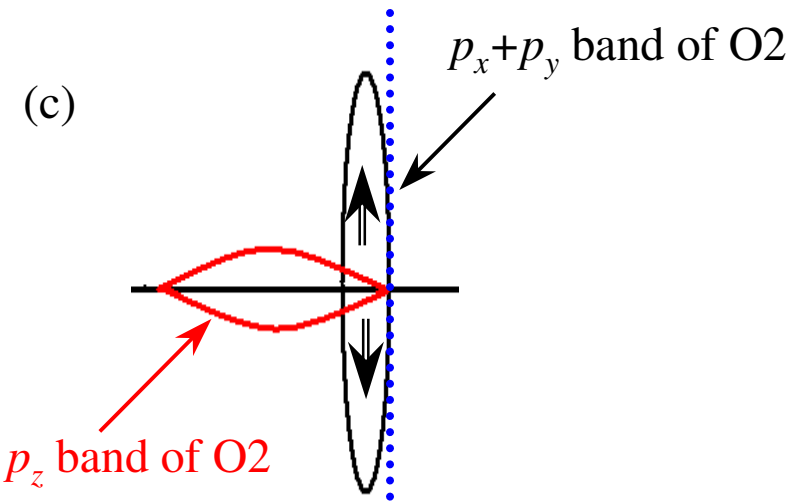
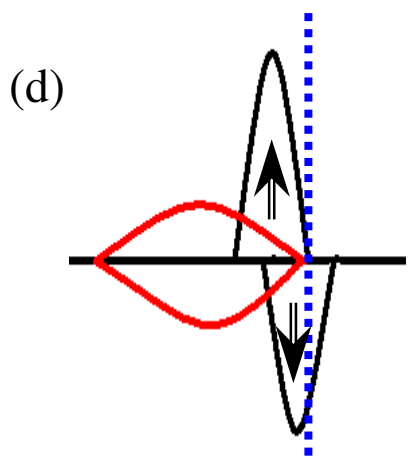
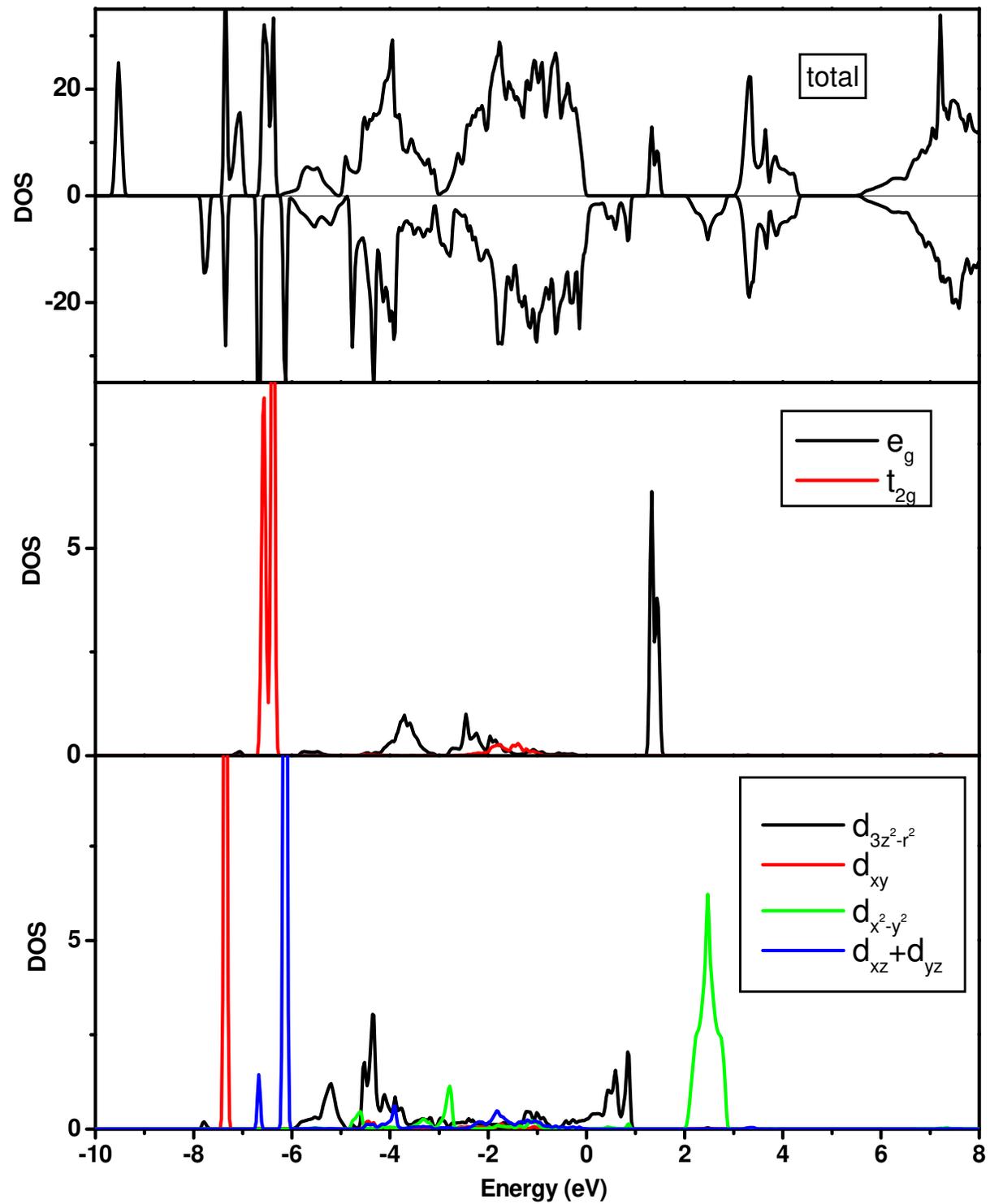
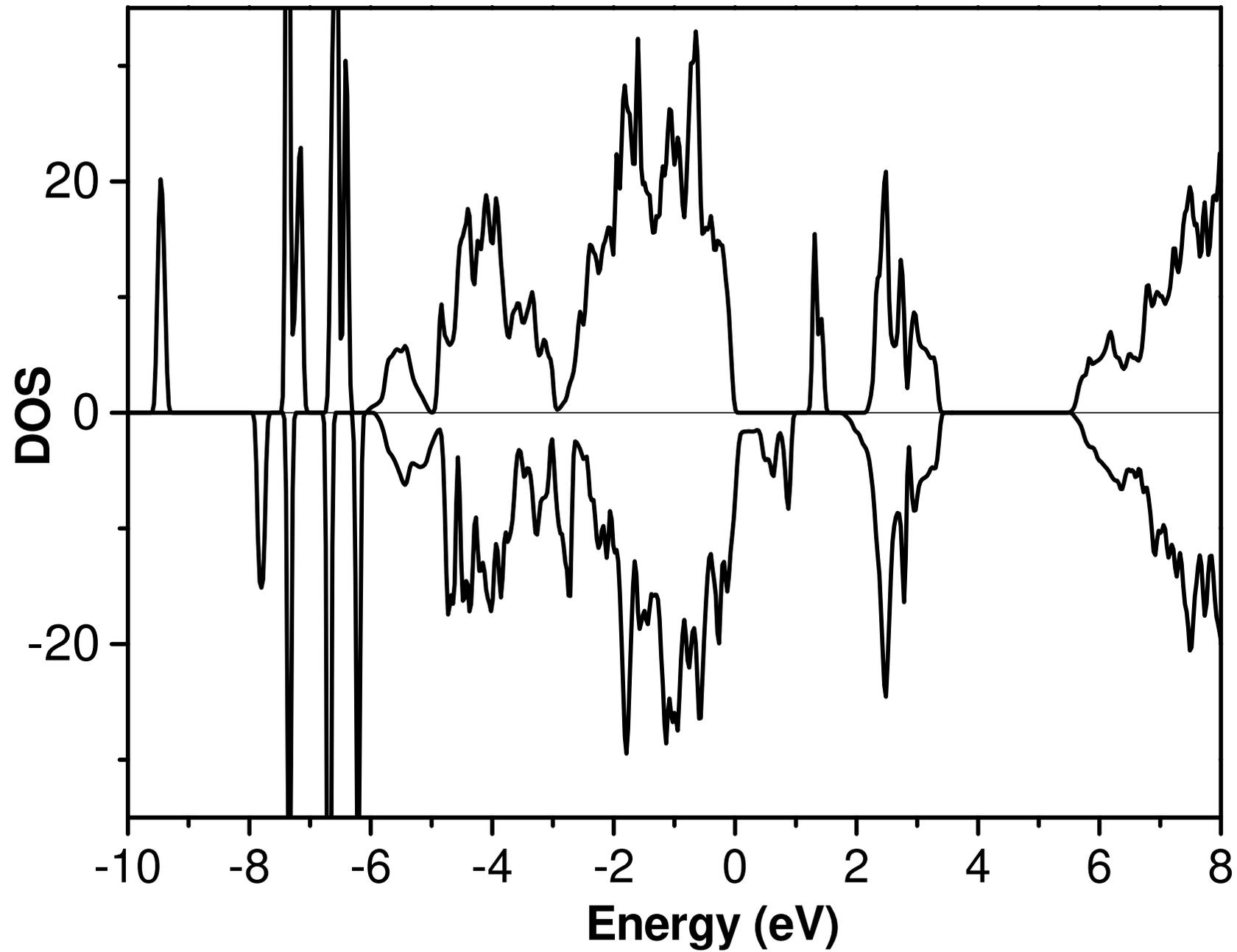
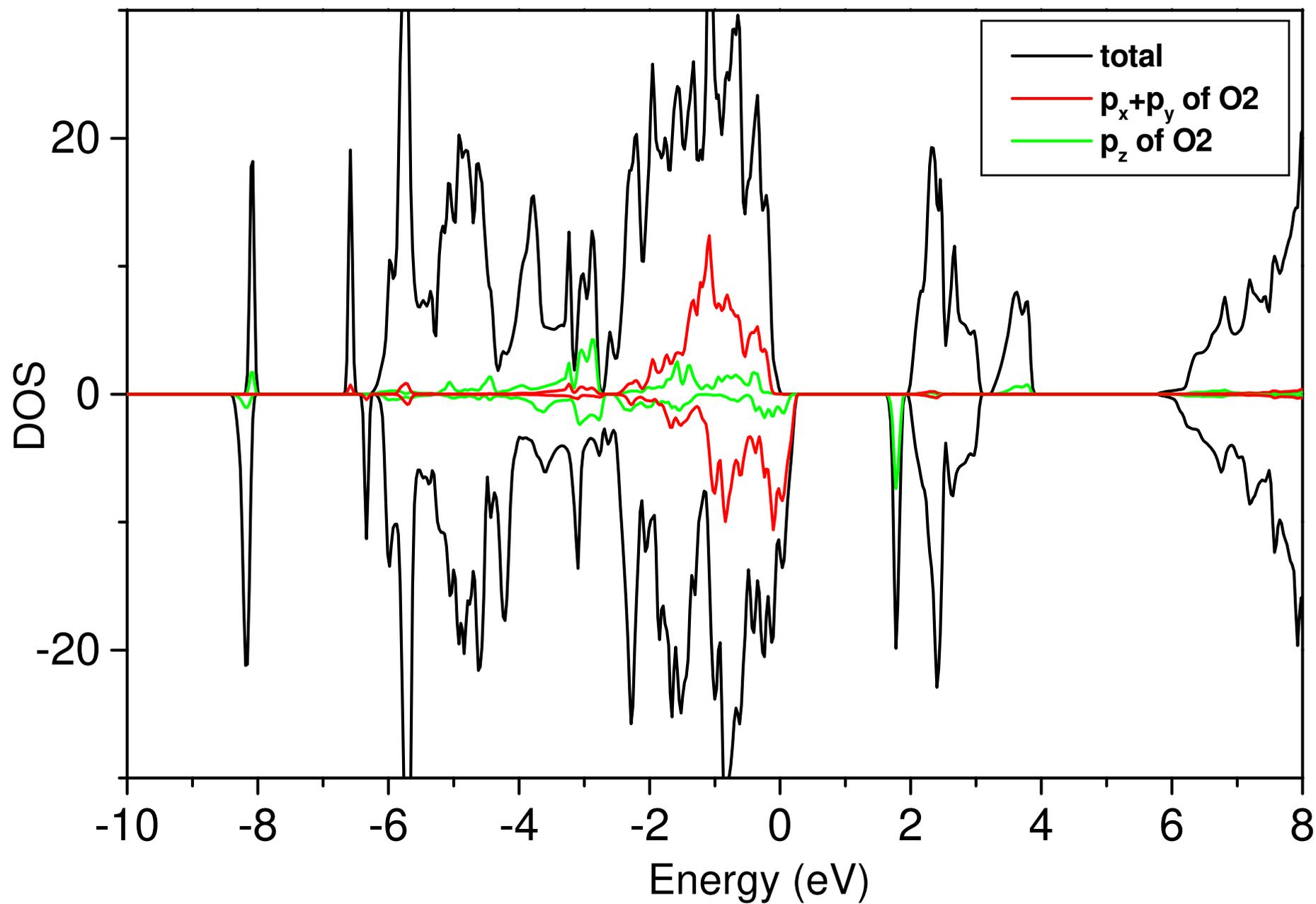
\end{document}